\documentstyle[11pt,epsfig]{article}
\newcommand{\be}{\begin{equation}}
\newcommand{\ee}{\end{equation}}
\newcommand{\ba}{\begin{eqnarray}}
\newcommand{\ea}{\end{eqnarray}}
\newcommand{\ket}[1]{| {#1} \rangle}
\newcommand{\bra}[1]{\langle {#1} |}
\newcommand{\ave}[1]{\langle {#1} \rangle}

\begin{document}

\begin{titlepage}
\vspace*{3.5cm}
\begin{center}
\begin{Large}
{\bf{A Non-Perturbative Treatment of the Pion in the Linear
Sigma-Model}\\}
\end{Large}
\vspace*{1.cm}
Z. Aouissat$^{(1)}$, G. Chanfray$^{(2)}$, P. Schuck$^{(3)}$
and J. Wambach$^{(1)}$,\\
$^{(1)}$ {\small{\it Institut f\"{u}r Kernphysik, Technische Hochschule 
Darmstadt, Schlo{\ss}gartenstra{\ss}e 9, D-64289 Darmstadt, Germany.}}\\
$^{(2)}$ {\small{\it IPN-Lyon, 43 Bd. du 11 Novembre 1918,
F-69622 Villeurbanne C\'{e}dex, France.}}\\
$^{(3)}$ {\small{\it  ISN, 53 avenue des Martyrs,
F-38026 Grenoble C\'edex, France.}}\\
\end{center}
\vspace*{2.cm}
\begin{abstract}
Using a non-perturbative method based on the selfconsistent 
Quasi-particle Random-Phase Approximation (QRPA) we describe the properties 
of the pion in the linear $\sigma$-model. It is found that the pion is 
massless in the chiral limit, both at zero- and finite temperature, in 
accordance with Goldstone's theorem.
\end{abstract}
\end{titlepage}
\newpage
\vspace*{1.5cm}
 
\section{Introduction}
Chiral Lagrangians, as the effective low-energy realization of QCD, have
become increasingly important in hadronic physics. In the sector of
up and down quarks and for vanishing quark masses, QCD exhibits
an exact global $SU(2)_L\times SU(2)_R$ chiral symmetry. In the 
non-perturbative vacuum, this symmetry is spontaneously broken to its 
vectorial subgroup $SU(2)_V$ with the appearence of pions as Goldstone
bosons. The interaction of the Goldstone bosons is greatly restricted by
chiral symmetry involving the ratios $(m_\pi/4\pi f_\pi)^2$ and
$(E_\pi/4\pi f_\pi)^2$ as a small expansion parameters, where $m_\pi$,
$E_\pi$ and $f_\pi$ denote the pion mass, the pion energy and the weak
pion decay constant respectively. A systematic expansion is provided  
by chiral perturbation theory \cite{GL1984}. For example, the $\pi-\pi$ 
scattering amplitude is determined order by order in the 
number of derivatives. In this way the low-energy theorems 
are known to be maintained. For many reasons, however, it would be
interesting to have a non-perturbative approach while still maintaining
the low-energy theorems. One obvious reason is the requirement of unitarity
of the S-matrix in the scattering problem. Another relates to the 
thermodynamics of effective chiral theories, especially in the study of 
chiral restauration. As the critical point is approached one cannot expect
perturbation theory to provide a valid description.
 
Needless to say that the question of preserving the symmetries 
non-perturbatively is a very delicate one \cite{ARCSW}. While in 
perturbative calculations the class of diagrams that ought to be 
considered in order to preserve the symmetries in the physical observables 
is well known, the situation is far less clear in the non-perturbative case.
The aim of the present paper is to demonstrate that such a program is indeed 
possible. Our theoretical framework will be the linear $\sigma$-model 
which is especially suited for the techniques to be employed. These 
techniques have their origin in many-body physics and consist of a  
mean-field treatment via a Bogoliubov rotation supplemented by RPA 
fluctuations.
It is well known that such an approach, while being non-perturbative,
treats symmetries and spontaneous symmetry breaking correctly \cite{Marsh}.
We shall demonstrate that, exactly as in the fermionic case, the RPA 
built on the selfconsistent mean field is able to restore the symmetry 
broken by the mean-field vacuum.
 
The paper is organized as follows: First the formulation of the 
bosonic mean-field problem will be given in sect.~2. In the 
'quasi-particle basis' thus obtained, the RPA excitation spectrum for the
single-pion mode is constructed in sect.~3. It will be shown explicitly 
that this spectrum contains a zero mode in the chiral limit, to be 
identified with the 'Goldstone pion'. In sect.~4 the formalism 
will be extended to finite temperature as a first step towards a 
non-pertubative description of the chiral phase transition. Again, there
is no mass generation in the chiral limit. Conclusions and an outlook
are given in sect.~5.

\section{The Bogoliubov Rotation}
The starting point is the Lagrangian density of the linear $\sigma$-model
\cite{GML60}
\\
\begin{equation}
 {\cal L}  =  \frac{1}{2}\left[ \left(\partial_{\mu}{\bf{\pi}}\right)^2
     + \left(\partial_{\mu}{\hat \sigma}\right)^2 \right]
 - \frac{\mu_{0}^2}{2} \left[ {\bf{\pi}}^2 + {\hat \sigma}^2 \right]
 - \frac{\lambda_{0}^2}{4}\left[ {\bf{\pi}}^2 + {\hat \sigma}^2 \right]^2
   + c {\hat \sigma}.
\label{eq1}
\end{equation}
\\
where $\lambda_0$ represents the bare coupling constant, $\mu_0$ the 
mass parameter and $\pi$ and ${\hat \sigma}$ denote the bare pion and sigma
fields, respectively. Chiral symmetry is explicitly broken (in the PCAC
sense) by the last term in the Lagrangian, $c {\hat \sigma}$. At tree level
the pion and sigma masses are given by
\\
\begin{eqnarray}
 m_{\pi}^2 &=& \mu_0^2 + \lambda_0^2  \ave{\hat \sigma}^2    
 \nonumber\\
 m_{\sigma}^2 &=& \mu_0^2 + 3\lambda_0^2  \ave{\hat \sigma}^2
\nonumber\\
 c &=& \ave{\hat \sigma} \mu_0^2 + \lambda_0^2  \ave{\hat \sigma}^3
\label{eq95}
\end{eqnarray}
\\
The pion possesses manifestly the Goldstone boson character since its mass 
is trivially proportional to $c$. The perturbative one loop calculation
preserves this result as is shown for instance in \cite{BWL69}.\\   
For further development it is now convenient to define the field operators
in terms of creation and annihilation operators as
\\
\begin{eqnarray}
  \pi_{i}({\bf x}) &=& \int \frac{d^3{\bf q}}
  {\sqrt{(2\pi)^{3} 2\omega_{q}}}
  \left(a_{{\bf q}\, i} e^{i{\bf q}{\bf x}} +a^{+}_{{\bf q}\, i}
  e^{-i{\bf q}{\bf x}} \right) \nonumber\\
  {\hat \sigma}({\bf x}) &=& \int \frac{d^3{\bf q}}
  {\sqrt{(2\pi)^{3} 2\omega_{q}}}
  \left(b_{{\bf q}} e^{i{\bf q}{\bf x}} + b^{+}_{{\bf q}}
  e^{-i{\bf q}{\bf x}} \right)
\label{eq2}
\end{eqnarray}
where the frequency $\omega_q$, common to both fields, is given by
\begin{equation}
      \omega_q = \sqrt{\mu_0^2 + q^2}.
\label{eq3}
\end{equation}

In a first step a canonical transformation is performed for the pion as 
well as the sigma field. Thus we introduce a new set of creation and 
annihilation operators through the following Bogoliubov rotation 
\\
\begin{eqnarray}
  \alpha ^{+}_{q} &=& u_{q}a^{+}_{q} - v_{q}a_{ q},
  \nonumber\\
  \beta ^{+}_{q} &=& x_{q}b^{+}_{q} -
    y_{q}b_{q} - w_q
\label{eq4}
\end{eqnarray}
\\
with $u_{q}$, $v_{q}$,  $x_{q}$ and $y_{q}$  being even functions of their 
argument, and $w_q$ a c-number. The first equation is the usual 
bosonic Bogoliubov transformation applied to the pion field. The new vacuum 
with respect to these $\alpha_q$ operators is the well-known 
'squeezed state'. In the second equation the transformation contains an
additional 'shift' $w_q$ to account for the macroscopic condensate 
$\ave{\hat \sigma}$. For later notation we will adopt the variable s to 
designate this condensate. To render the transformations canonical the 
Bogoliubov factors have to obey the following constraints
\\
\begin{equation}
       u_{q}^{2} -  v_{q}^{2} = 1, \quad\quad\quad
       x_{q}^{2} -  y_{q}^{2} =1.
\label{eq5}
\end{equation}
\\
In the 'quasi-boson' basis eq.(\ref {eq4}) the fields and their conjugates read
\\
 \begin{eqnarray}
  \pi_{j}({\bf x}) &=&  \int \frac{d^3{\bf q}}
  {\sqrt{(2\pi)^{3} 2\omega_{q}}}
   \left( u_{q}+v_{q} \right)
  \left(\alpha_{q\, j} e^{i{\bf q}{\bf x}} +\alpha^{+}_{q\, j}
  e^{-i{\bf q}{\bf x}} \right) \nonumber\\
 {\dot \pi}_{j}({\bf x}) &=&
 \int \frac{d^3{\bf q}}
  {\sqrt{(2\pi)^{3} 2\omega_{q}}}
   (-i \omega_{q}) \left( u_{q}-v_{q} \right)
  \left(\alpha_{q\, j} e^{i{\bf q}{\bf x}} - \alpha^{+}_{q\, j}
  e^{-i{\bf q}{\bf x}} \right)\nonumber\\
  \sigma({\bf x}) &=&  \int \frac{d^3{\bf q}}
  {\sqrt{(2\pi)^{3} 2\omega_{q}}}
   \left( x_{q} + y_{q} \right)
  \left(\beta_{q} e^{i{\bf q}{\bf x}} +\beta^{+}_{q}
  e^{-i{\bf q}{\bf x}} \right) \nonumber\\
 {\dot \sigma}({\bf x}) &=& \int \frac{d^3{\bf q}}
  {\sqrt{(2\pi)^{3} 2\omega_{q}}}
   (-i \omega_{q}) \left( x_{q}-y_{q} \right)
  \left(\beta_{q} e^{i{\bf q}{\bf x}} - \beta^{+}_{q}
  e^{-i{\bf q}{\bf x}} \right)
 \label{eq6}
 \end{eqnarray}
\\
and the quasi-boson vacuum  $\ket{ \Phi}$ 
( $\alpha  \ket{ \Phi}= \beta  \ket{ \Phi} = 0$ )
is given by the following coherent state
\\
\begin{equation}
\ket{ \Phi} = exp\left[ \sum_{q} z_1(q)a_q^+a_{-q}^+ \, + \, z_2(q) b_q^+b_{-q}^+\, +\, w_q b_{-q}^+ \right] \ket{ 0}.
\label{eq7}
\end{equation}
\\
were $\ket{ 0 }$ denotes the vacuum for the original basis 
($ a_q\ket{ 0 } =b_q \ket{ 0 } =0$ )
and  $ z_1 = \frac{v}{u},\quad z_2= \frac{y}{x}$.\\
It is now straightforward to write the Hamiltonian of the linear sigma
model in the quasi-particle basis. After normal-ordering one obtains
\\
 \begin{eqnarray}
    H &=& {\cal H}_{0}(v,y,s) \,+\,
      \eta \left[ \beta_0 + \beta_0^+ \right]
   \,\,+\,\,
  \sum_{q}
    {\cal E}_{\pi}(q) \left[ \alpha_{j}^{+}(q) \alpha_{j}(q)
    \right] \,\,+\,\,
  \sum_{q}
    {\cal E}_{\sigma}(q) \left[ \beta^{+}(q) \beta(q)
    \right]
    \nonumber\\
 &+&
  \sum_{q}
     c_{\pi}(q) \left[ \alpha_{j}^{+}(q) \alpha_{j}^{+}(-q) +
   \alpha_{j}(q) \alpha_{j}(-q) \right]
 \,\,+\,\,
  \sum_{q}
     c_{\sigma}(q) \left[ \beta^{+}(q) \beta^{+}(-q) +
   \beta(q) \beta(-q) \right]  \nonumber\\
    &+&
  \int d{\bf x} \,:\,\, \left[ \lambda_0^2 s \sigma({\bf x})
  \left( {\bf{\pi^2(x)}} + {\sigma^2(x)} \right) \,+\,
 \frac{\lambda_0^2}{4}\left( {\bf{\pi^2(x)}} + {\sigma^2(x)} \right)^2
 \right] \,\,:
\label{eq8}
 \end{eqnarray}
\\
where ":" in the interaction part of $H$ denotes normal ordering (to
avoid lengthy expressions the interaction part is given in terms of
field operators rather than in second-quantized form).

The pion and sigma fields are given by
eq.(\ref{eq6}), and the coefficients  ${\cal H}_{0}$,
$\eta$, ${\cal E}_{\sigma \,,\, \pi}$ and $c_{\sigma\,,\,\pi}$ read 
explicitly
\\
\begin{eqnarray}
 {\cal H}_{0}(v,y,s) &=& \sum_{q}\,
 \omega_{q} ( 3 v^2_{q} +y^2_{q}+2 )
 \,+\, \frac{3 \lambda_0^2}{4} \left[ J^2_0 +5 I^2_0 + 2I_0J_0 \right]
 \,+\, \frac{3 \lambda_0^2 s^2}{2}\left[ I_0 + J_0 \right]
 \,+\, \frac{\mu_0^2 s^2}{2} + \frac{\lambda_0^2 s^4}{4} -cs
 \nonumber\\
\eta &=& \frac{x_{0} +y_{0}}{\sqrt{\mu}}
\left[ 3\lambda_0^2 s I_0 + 3\lambda_0^2 s J_0 + \lambda_0^2 s^3 + \mu_0^2 s -  c
\right]
\nonumber\\
   c_{\pi}(q) &=&  \omega_{q} (u_{q}v_{q}) +
   \frac{\lambda_0^2}{2}
   \frac{(u_{q}+v_{q})^{2}}{2\omega_{q}}
    \left[ 5I_0 + J_0 +  s^2 \right]
 \nonumber\\
   c_{\sigma}(q) &=&  \omega_{q} (x_{q}y_{q}) +
 \frac{3 \lambda_0^2}{2}
   \frac{(x_{q}+y_{q})^{2}}{2\omega_{q}}
    \left[ I_0 + J_0 +  s^2 \right]
\nonumber\\
 {\cal E}_{\pi}(q) &=&  \omega_{q} (u_{q}^2 + v_{q}^2) +
   \lambda_0^2
   \frac{(u_{q}+v_{q})^{2}}{2\omega_{q}}
    \left[ 5I_0 + J_0 +  s^2 \right]
 \nonumber\\
 {\cal E}_{\sigma}(q) &=&  \omega_{q} (x_{q}^2 + y_{q}^2) +
 3 \lambda_0^2
   \frac{(x_{q}+y_{q})^{2}}{2\omega_{q}}
    \left[ I_0 + J_0 +  s^2 \right]
\label{eq9}
\end{eqnarray}
\\
Here $I_0$ and  $J_0$ are quadratically divergent integrals 
\\
 \begin{equation}
  I_0  =  \int \frac{d^{3}\vec{q}}{(2\pi)^{3}}
  \quad \frac{(u_{q}+v_{q})^2}{2\omega_{q}}, \quad \quad \quad
   J_0 = \int \frac{d^{3}\vec{q}}{(2\pi)^{3}}
  \quad \frac{(x_{q}+y_{q})^2}{2\omega_{q}}.
\label{eq10}
 \end{equation}
arising from the tadpole loops in the selfenergies (see Fig.~1).
\\
As usual the amplitudes $u_i$ , $v_i$ and s are determined by minimizing
the vacuum expectation value $\frac{\bra{\Phi } H \ket{ \Phi }}
{\ave{\Phi\mid \Phi} }$. 
This is in fact equivalent to demanding that the single-particle part
of H be diagonal {\sl i.e.} $c_{\pi, \sigma} =0 $, and that the term linear 
in the boson operators vanishes, {\sl i.e.} $\eta = 0 $.
Defined in this way the set of $\alpha$ and $\beta$ operators form  
the 'selfconsistent quasiparticle basis' (scqb).\\
We now turn to the evaluation of the amplitudes $u$, $v$ and
$x$ and $y$. First we note that the
expressions for 
 $c_{\pi, \sigma}$ and ${\cal E}_{\pi, \sigma}$ can be recast in the form
\\
\begin{eqnarray}
   c_{\Phi}(q) &=&  (U_{q}V_{q}) e_{\Phi}(q)
   + \frac{(U_{q}^{2}+ V_{q}^{2})}{2}
    \Delta_{\Phi}(q) = 0   \nonumber\\
    \nonumber\\
   {\cal E}_{\Phi}(q) &=&  (U_{q}^{2}+V_{q}^{2}) e_{\Phi}(q)
    + (2U_{q}V_{q})
    \Delta_{\Phi}(q)
\label{eq11}
\end{eqnarray}
\\
For notational purposes a generic field $\Phi$ has been introduced to
designate either the pion or the sigma, and the corresponding
Bogoliubov parameters (U,V) denote the pair (u,v) or (x,y). The following 
identities are easily verified
\\
\begin{eqnarray}
   e_{\pi}(q) &=&  \omega_{q} \,+\, \Delta_{\pi}(q),
  \quad\quad\quad
  \Delta_{\pi}(q) = \frac{\lambda_0^2}
  {2\omega_{q}}\left[5 I_0 + J_0 + s^2 \right]
    \nonumber\\
   e_{\sigma}(q) &=&  \omega_{q} \,+\, \Delta_{\sigma}(q),
  \quad\quad\quad
  \Delta_{\sigma}(q) = \frac{3\lambda_0^2}
  {2\omega_{q}}\left[I_0 + J_0 + s^2 \right].
\label{eq12}
\end{eqnarray}
\\
With the above expressions and some trivial algebra one can extract 
selfconsistently the Bogoliubov factors from
\\
\begin{equation}
   (U_{q}+V_{q})^2 =  \frac{e_{\Phi}(q) - \Delta_{\Phi}(q)}
   {\sqrt{e_{\Phi}^2(q) - \Delta_{\Phi}^2(q)}},
\label{eq13}
\end{equation}
\\
and the quasiparticle energies are given by
\\
\begin{equation}
 {\cal E}_{\Phi}(q) \, = \, (U_{q} - V_{q})^2 \omega_{q}
 \, =\, \sqrt{q^2 + {\cal E}_{\Phi}^2(0)}
\label{eq14}
\end{equation}
\\
This result allows to reexpress the BCS gap equations for the auxiliary 
variables (U,V) in terms of more physical variables namely the quasi-pion 
and quasi-sigma masses as
\\
\begin{eqnarray}
 {\cal E}_{\pi}^2(0)&=& \mu_0^2 + \lambda_0^2 \left[5I_0 +J_0 +s^2\right]
 \nonumber\\
 {\cal E}_{\sigma}^2(0)&=& \mu_0^2 + 3\lambda_0^2 \left[I_0 +J_0 +s^2\right]
\label{eq15}
\end{eqnarray}
\\
In order to derive the BCS equations one should recall that we have made 
use of the two conditions $c_{\pi} = c_{\sigma} = 0$ arising from the 
minimization of ${\cal H}_0(v,y,s)$ with respect to v and y. The 
minimization with respect to s, on the other hand, yields an additional 
condition, namely $\eta = 0$. This will fix the shift s via 
\\
\begin{equation}
   \mu_0^{2} + \lambda_0^{2} s^{2}  +
  3 \lambda_0^{2} \left[ I_0 +J_0 \right] = \frac{c}{s}.
 \label{eq16}
 \end{equation}
\\
The HFB results given above can be summarized diagrammatically as 
indicated in Fig.~1 \\
\noindent\begin{figure}[hbt]
\centerline{ 
\epsfig{file=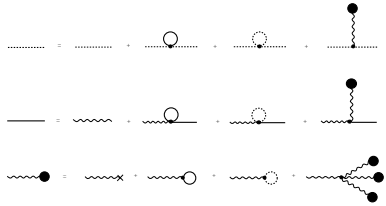,height=5cm,angle=270}}
\caption[fig1]{\small
Diagrammatic representation of the  mean-field results
for the pion and sigma. The dashed line denotes the selfconsistent quasi pion
propagator, the solid line the quasi sigma, and the wavy line the two point Greens function of the bare field ${\hat \sigma}$ of the Lagrangian density in 
eq.(\ref{eq1}). 
\label{fig1} }
\end{figure}
\\
For a physical interpretation it is important to see now how the 
quasiparticle masses behave in the chiral limit ($c\to 0$). 
With the expressions given above the latter can be written as
\\
\begin{eqnarray}
 {\cal E}_{\pi}^2(0)&=& \frac{c}{s}
 + 2\lambda_0^2 \left[I_0 -J_0 \right],
 \nonumber\\
 {\cal E}_{\sigma}^2(0)&=& \frac{c}{s} +2\lambda_0^2 s^2.
\label{eq17}
\end{eqnarray}
\\
which implies that the quasipion mass does not vanish due to 
the nonvanishing difference $ I_0 -J_0$. This is in violation of Goldstone's
theorem. To restore the symmetry one has to go further and this will be done
in the next section.\\
Before ending we wish to comment on the difference of the quasi-particle
masses which can be written as 
\\
\begin{equation}
 {\cal E}_{\sigma}^2(0)-{\cal E}_{\pi}^2(0)= \frac{ 2\lambda_0^2 s^2}
{ 1\quad-\quad 2 \lambda_0^2 \Sigma_{\pi\sigma}(0)}
\label{eq33}
\end{equation}
\\
and which arises from the finite value of the condensate, s, in the 
Goldstone phase. The explicit form of $\Sigma_{\pi\sigma}(0)$ will be given later on.
 The expression eq.(\ref{eq33}) is reminiscent of a Ward identity which links the 
three point 
function or  $\pi\pi\sigma$ vertex to the mass difference. An interesting 
feature of the identity above is that it only contains a 
weak divergence. By a simple redefinition of the coupling constant 
\\
\begin{equation}
  \lambda_0^2 = \frac{\lambda^2}{1+2\lambda^2L_0}
\label{eq34}
\end{equation}
\\
($L_0$ diverges only logarithmically) it can be rendered finite. 
It will turn out later that this redefinition of the bare coupling constant 
is also able to make the RPA solutions free of divergences. The full 
renormalization program will be discussed in a forthcoming paper.

\section{Single-Pion RPA}
Given the mean-field results presented in the last section the task is 
to obtain a Goldstone mode in the chiral limit. As is well known in 
many-body physics the restoration of a symmetry, which is broken at 
mean-field level, is provided by the ``selfconsistent'' RPA. To make 
this explicit for the case at 
hand we remind that $Q_5^a \ket{ vac }$ represents the single-pion mode, 
where $Q_5^a$ is the axial charge given by the volume
integral of the time component of the axial vector current. In the 
linear sigma model the current is given by 
\\
\begin{equation}
 A_{5}^{a \, \mu} =
 \sigma
 ( \partial^{\mu}\pi^{a} )
 - ( \partial^{\mu} \sigma)
 \pi^{a}.
\label{eq18}
\end{equation}
\\
When expressed in the selfconsistent quasiparticle basis the axial charge 
then becomes 
\\
\begin{eqnarray}
 Q_{5}^{a} = \sqrt{\frac{(2\pi)^3}{2{\cal E}_{\pi}(0)}}\,
  i{\cal E}_{\pi}(0) &s&
 \left[ \alpha^{a\, +}_0\, -\, \alpha^a_0 \right] \quad+\quad
 \sum_{q} i \frac {{\cal E}_{\pi}(q)-{\cal E}_{\sigma}(q)}
 {\sqrt{4{\cal E}_{\pi}(q) {\cal E}_{\sigma}(q)}}
  \left[\beta^+_q \alpha^{a\, +}_{-q}
  \,-\,  \beta_{-q} \alpha^{a}_q \right] \nonumber\\
  &+&\quad\quad
 \sum_{q} i \frac {{\cal E}_{\pi}(q)+{\cal E}_{\sigma}(q)}
 {\sqrt{4{\cal E}_{\pi}(q) {\cal E}_{\sigma}(q)}}
  \left[\beta_q \alpha^{a\, +}_{q}
  \,-\,  \beta^+_{-q} \alpha^{a}_{-q} \right]
\label{eq19}
\end{eqnarray}
\\
A remark is in order. We see that the operator $Q_5^a$, when acting on the 
coherent state $\ket{ \Phi}$, as defined in eq.~(\ref{eq7}),
can excite six different modes corresponding to a single-pion excitation 
and pairs of correlated pion and sigma excitations. When written in the
original basis $Q_5^a$ takes the same form as in eq.~(\ref{eq19}) except 
that the quasi-particle masses ${\cal E}_{\pi}, {\cal E}_{\sigma}$ are
replaced by the tree-level masses $m_{\pi}, m_{\sigma}$ (eq.~(\ref{eq95})).
It therefore leads to the same modes. The quasi-particle basis has the 
advantage, however, that in the chiral limit $(c=0)$ all modes survive, 
while in the original basis the single-pion excitation vanishes, since the 
tree-level pion mass goes to zero in that limit. We will see below that
the quasiparticle representation is indeed needed.\\
To proceed further, we consider the following  RPA excitation
operator $Q_{\nu}^+$
\\
\begin{equation}
 Q_{\nu}^+ =
 X^1_{\nu} \alpha^{a\, +}_0 \,\, -\,\,  Y^1_{\nu} \alpha^a_0
 \quad+\quad
 \sum_{q}
  \left[ X_{\nu}^2(q)  \beta^+_q \alpha^{a\, +}_{-q}
 \, \,-\,\,  Y_{\nu}^2(q)  \beta_{-q} \alpha^{a}_q \right]
\label{eq20}.
\end{equation}
\\
As usual the RPA ground-state correlations will be determined by the
requirement that $Q_{\nu}\ket{ RPA} = 0$. Applying the equation
of motion method of Rowe \cite{Row68, RiSch}, one then has
\\
\begin{equation}
 \bra{RPA } \left[ \delta Q_{\nu} \, \, , \, \, \left[H\, , \, Q_{\nu}^+
 \right] \right] \ket{ RPA} = \Omega_{\nu}
 \bra{RPA } \left[ \delta Q_{\nu}  \, , \, Q_{\nu}^+
 \right] \ket{ RPA}
\label{eq21}
\end{equation}
\\
which leads to the RPA equations for boons. In case of an exact symmetry, one 
particular solution has to be 'spurious', {\sl i.e.} occurs at zero energy 
($\Omega_{\nu}=0$). 
To identify this solution one has to consider the operator which generates 
the symmetry. It has to be ensured, of course, that the latter
possesses the excitations that are present in the general
ansatz of the RPA operator $Q^+_{\nu}$. Indeed one notices that the
chiral symmetry operator $Q_5^a$, when written in the original basis, has
the same structure as the RPA operator. Two difficulties occur, however.
The first has been eluded to and relates to the disappearance of the 
single-pion component from the symmetry operator when going to the chiral 
limit.  The second difficulty is caused by the presence of the 
'mixed' combinations
$b_q a^{a\,+}_q$ and $b^+_{-q} a_{-q}^a$ in $Q_5^a$. Such terms are
undesirable since $Q_5^a$ is no longer a solution of eq.(\ref{eq21}). 
When written in the quasi-particle basis, the first problem is automatically
cured, as mentioned above. The second, at first glance, seems to persist 
since the 'mixed' terms are still present. These terms give no contributions 
to the RPA equations , however, as long as the Hamiltonian is diagonal.
By construction this is the case, of course. \\
To make the spurious solution explicit, we consider
the set of 4 coupled equations resulting from the explicit form of the
RPA excitation operator $Q^+_\nu$ in eq.~(\ref{eq20}). 
Using Feshbach projection techniques it is
advantageous to first solve the scattering problem for the pair of 
quasi-sigma and pion (lower part of Fig.~2) which is generated
by the last two terms in $Q^+_\nu$. 
In the single-pion subspace one then has to solve a Dyson equation (upper
part of Fig.~2) to finally obtain the physical pion mass.\\
\noindent\begin{figure}[hbt]
\centerline{ 
\epsfig{file=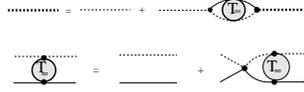,height=4cm,angle=270}}
\caption[fig2]{\small
Upper part:  The Dyson equation for the physical pion (thick dashed lines) 
for which the mass operator has been extracted from the scattering of the 
quasibosons in an RPA equation.
Lower part: The scattering equation for a pair of quasi-sigma 
(thin full lines) and quasi-pion (thin dashed lines). 
\label{fig2} }
\end{figure}
\\
This can in fact be done analytically and yields
\\
\begin{equation}
\Omega^2_{\nu}\,=\, {\cal E}^2_{\pi}(0) \quad+\quad
\frac{ 4 \lambda_0^4 s^2 \Sigma_{\pi\sigma}(\Omega^2_{\nu})}
{ 1\,\,-\,\, 2 \lambda_0^2 \Sigma_{\pi\sigma}(\Omega^2_{\nu})}
\label{eq22}
\end{equation}
\\
where $\Sigma_{\pi\sigma}$ is the contribution of the (quasi) pion-sigma 
bubble to the pion selfenergy given by
\\
\begin{equation}
 \Sigma_{\pi\sigma}(\Omega^2_{\nu}) \,=\,
 \int \frac{d^{3}\vec{q}}{(2\pi)^{3}} \quad
 \frac {{\cal E}_{\pi}(q)+{\cal E}_{\sigma}(q)}
 {2{\cal E}_{\pi}(q) {\cal E}_{\sigma}(q)}
 \frac {1}{\Omega^2_{\nu}\, -\,
 ({\cal E}_{\pi}(q) + {\cal E}_{\sigma}(q))^2 } .
\label{eq23}
\end{equation}
\\
Using the identity
\\
\begin{equation}
 \Sigma_{\pi\sigma}(0) \,=\,
 \frac{I_0 \,\, -\,\, J_0}
  {{\cal E}^2_{\pi}(0)-{\cal E}^2_{\sigma}(0)},
\label{eq24}
\end{equation}
\\
we obtain after some algebra, 
\\
\begin{equation}
\Omega^2_{\nu}\,=\,
\frac{ 2\lambda_0^2 \left[{\cal E}_{\pi}^2(0) \,-\,
 {\cal E}_{\sigma}^2(0) \right]
\left[\Sigma_{\pi\sigma}(0)
 \,-\, \Sigma_{\pi\sigma}(\Omega^2_{\nu})\right]}
{ 1\quad-\quad 2 \lambda_0^2 \Sigma_{\pi\sigma}(\Omega^2_{\nu})}
\quad+\quad
\frac{c}{s}\ .
\label{eq25}
\end{equation}
\\
In the chiral limit $(c=0)$ the zero-energy solution is now manifest 
(independent of any regularization scheme).\\ 
It is interesting to ask: What is the Goldstone boson dispersion relation?
{\sl i.e.} what is the behavior of the spurious mode under spatial 
translations.
For this purpose we shall use as the generator of the spurious mode the time
component of the axial vector current $A^{a\, 0}_5(x)$ rather than the 
axial charge. The Fourier transform of the latter allows
to pick up the spurious mode at any finite three momentum. To show that 
$A^{a\,0}_5(x)$ can generate a spurious mode we recall the PCAC relation
\\
\begin{equation}
 \partial_{\mu} A_{5}^{a \, \mu}({\bf x}) =c \pi^a({\bf x}).
\label{eq26}
\end{equation}
\\
Using Heisenberg's equation of motion, after Fourier transformation,
this can be simply expressed as
\\
\begin{equation}
 \left[H, A_5^{a\,0}({\vec p})\right]  \,\,+\,\,
   {\vec p} {\vec A}_5^a({\vec p}) \,\,=\,\, -ic \pi^a({\vec p})
\label{eq27}
\end{equation}
\\
In the chiral limit the single-pion part of the RPA operator then generates
a solution of finite three-momentum which has the following property
\\
\begin{equation}
 \bra{RPA } \left[ \delta Q_{\nu} \, \, , \, \,
 \left[H, A_5^{a\,0}({\vec p})
 \right] \right] \ket{ RPA }  \,\,+\,\,
 {\vec p} \bra{RPA } \left[ \delta Q_{\nu}  \, , \,
 {\vec A}_5^a({\vec p})
 \right] \ket{ RPA} \,\,=\,\, 0.
\label{eq28}
\end{equation}
\\
This clearly indicates that for pions at rest $({\vec p} ={\vec 0})$
again a zero-frequency solution exist. To make the 
dispersion relation explicit we consider the following excitation 
operator
\\
\begin{equation}
 Q_{\nu}^+({\vec p}) =
 X^1_{\nu}({\vec p}) \alpha^{a\, +}_{- \vec p}
  \,\, -\,\,  Y^1_{\nu}({\vec p}) \alpha^a_{\vec p}
 \quad+\quad
 \sum_{q}
  \left[ X_{\nu}^2({\vec p},{\vec q})
  \beta^+_{{\vec q}-{\vec p}} \alpha^{a\, +}_{-{\vec q}}
 \, \,-\,\,  Y_{\nu}^2({\vec p}, {\vec q})
  \beta_{{\vec p}-{\vec q}} \alpha^{a}_{\vec q} \right]
\label{eq29}
\end{equation}
\\
which is the extension of eq.~(\ref{eq20}) to finite three-momenta.
After some algebra the RPA frequencies can be expressed as
\\
\begin{equation}
\Omega^2_{\nu}(\vec{p})\,=\,
\frac{ 2\lambda^2 \left[{\cal E}_{\pi}^2(0) \,-\,
   {\cal E}_{\sigma}^2(0) \right]
\left[\Sigma_{\pi\sigma}(0)
 \,-\, \Sigma_{\pi\sigma}(\Omega^2_{\nu}, {\vec p})\right]}
{ 1\quad-\quad 2 \lambda^2 \Sigma_{\pi\sigma}(\Omega^2_{\nu}, {\vec p})}
\quad+\quad
\frac{c}{s} \quad+\quad  {\vec p}^2
\label{eq30}
\end{equation}
\\
with
\\
\begin{equation}
 \Sigma_{\pi\sigma}(\Omega^2_{\nu}, {\vec p}) \,=\,
 \int \frac{d^{3}\vec{q}}{(2\pi)^{3}} \quad
 \frac {{\cal E}_{\pi}({\vec q})+{\cal E}_{\sigma}({\vec p}-{\vec q})}
 {2{\cal E}_{\pi}({\vec q}) {\cal E}_{\sigma}({\vec p}-{\vec q})}
 \frac {1}{\Omega^2_{\nu}\, -\,
 ({\cal E}_{\pi}({\vec q}) + {\cal E}_{\sigma}({\vec p}-{\vec q}))^2 }.
\label{eq31}
\end{equation}
\\
Using the fact that the  $\pi\sigma$ selfenergy is Lorentz invariant one 
easily verifies that
\\
\begin{equation}
\Omega^2_{\nu}(\vec{p})\,=\,\Omega^2_{\nu}(0)  \quad+\quad  {\vec p}^2
\label{eq32}
\end{equation}
as it should be. Note that this result remains valid away from the
chiral limit.
\\
\section{Finite Temperature HFB-RPA}
In a first step towards a nonperturbative description of the chiral
phase transition we now extend the formalism to finite temperature
by using well-known methods available in the 
literature\cite{Goo80,Som83,TST85}. 
As we have demonstrated in the previous sections, the HFB-RPA has proven 
successful in preserving the symmetry which is manifest through the presence
of the spurious mode in the single pion RPA spectrum. The same is to be
expected at finite temperature. 
Let us now first come to the mean field problem.  We recall therefore that 
the thermodynamics of a gas of pions and sigmas at given 
temperature $T$ is governed by the free
energy $\Omega$ 
\\
\begin{equation}
 \Omega  =  \ave{H}  - TS.
\label{eq47}          
\end{equation}
\\
where $\ave{H}$ is the thermal expectation value of the Hamiltonian in 
eq.~(\ref{eq8}) and $S$ denotes the entropy. In thermal equilibrium the 
distribution of maximum entropy is the one which minimizes $\Omega$. 
The entropy then  reads
\\
\begin{equation}
 S= k_B \sum_{\nu}\left[(1+f_{\nu})ln(1+f_{\nu})-f_{\nu}lnf_{\nu}\right] 
\label{eq48}
\end{equation}
\\    
where $k_B$ is the Boltzmann constant and $f_{\nu}$ is the usual bosonic 
distribution functions. The sum $\nu$ includes the number of 
species as well as the three momentum q. 

In analogy to the zero-temperature case we can perform a 
temperature-dependent Bogoliubov rotation for both the $\pi$ and $\sigma$ 
fields. By making use of the Bloch-De Dominicis theorem \cite{BlDo} 
normal ordering on the rotated creation and annihilation operators 
can be carried out and one straightforwardly arrives at the HFB expression
for the free energy 
\\
\begin{eqnarray}
 \Omega  &\,=\,& {\cal H}_{0}(v_T,y_T,s_T) - TS,\nonumber\\
 {\cal H}_{0}(v_T,y_T,s_T)
  &\,=\,& \sum_{q}\,  \omega_{q} ( 3 v^2_{T,q} + y^2_{T,q}+2 )
 \,+\, \frac{3 \lambda_0^2}{4} \left[ J^2_T +5 I^2_T + 2I_TJ_T \right]
\nonumber\\
 &\,+\,& \frac{3 \lambda_0^2 s_T^2}{2}\left[ I_T + J_T \right]
 \,+\, \frac{\mu_0^2 s_T^2}{2} + \frac{\lambda_0^2 s_T^4}{4} -cs_T
\label{eq49}
\end{eqnarray}
\\
where ${\cal H}_{0}(v_T,y_T,s_T)$ is just the expectation value of H 
on the grand canonical ensemble. Minimizing $\Omega$ with respect to 
$u_{T,q}$ and $x_{T,q}$ and $s_T$ while keeping the canonical 
normalization of the Bogoliubov factors as in  eq.~(\ref{eq5}) then leads 
to the following identities\\
\\
\begin{eqnarray}
\eta^T &=& \frac{x_{T,0} +y_{T,0}}{\sqrt{\mu}}
\left[ 3\lambda_0^2 s_T I_T + 3\lambda_0^2 s_T J_T + \lambda_0^2 s_T^3 + \mu_0^2 s_T -  c \right]\,=\,0
\nonumber\\
   c^T_{\pi}(q) &=&  \omega_{q} (u_{T,q}v_{T,q}) +
   \frac{\lambda_0^2}{2}
   \frac{(u_{T,q}+v_{T,q})^{2}}{2\omega_{q}}
    \left[ 5I_T + J_T +  s_T^2 \right]\,\,=\,\,0 
\nonumber\\
   c^T_{\sigma}(q) &=&  \omega_{q} (x_{T,q}y_{T,q}) +
 \frac{3 \lambda_0^2}{2}
   \frac{(x_{T,q}+y_{T,q})^{2}}{2\omega_{q}}
    \left[ I_T + J_T +  s_T^2 \right]\,\,=\,\,0
\nonumber\\
 {\cal E}^T_{\pi}(q) &=&  \omega_{q} (u_{T,q}^2 + v_{T,q}^2) +
   \lambda_0^2
   \frac{(u_{T,q}+v_{T,q})^{2}}{2\omega_{q}}
    \left[ 5I_T + J_T +  s_T^2 \right]
 \nonumber\\
 {\cal E}^T_{\sigma}(q) &=&  \omega_{q} (x_{T,q}^2 + y_{T,q}^2) +
 3 \lambda_0^2
   \frac{(x_{T,q}+y_{T,q})^{2}}{2\omega_{q}}
    \left[ I_T + J_T +  s_T^2 \right]
\label{eq50}
\end{eqnarray}
\\
where the definitions are as in the zero-temperature case and 
${\cal H}_0$ takes the same form as in eq.~(\ref{eq9}).  
The loop integrals $I_T$ and  $J_T$ are now given by
\\
 \begin{equation}
   I_T =  \int \frac{d^{3}\vec{q}}{(2\pi)^{3}}
  \quad \frac{1+2f_{\pi}(q)}{2\omega_{q}(u_{T,q}-v_{T,q})^2}, \quad \quad \quad
   J_T = \int \frac{d^{3}\vec{q}}{(2\pi)^{3}}
  \quad \frac{1+2f_{\sigma}(q)}{2\omega_{q}(x_{T,q}-y_{T,q})^2}
\label{eq51}
 \end{equation}
\\
and the quasi-particle masses ${\cal E}^T_{\pi}$, ${\cal E}^T_{\sigma}$ and 
the condensate  $s_T$ take the form
\\
\begin{eqnarray}
 {\cal E}_{\pi}^T(0)^2&=& \mu_0^2 + \lambda_0^2 \left[5I_T +J_T +s_T^2\right] 
 \nonumber\\
 {\cal E}_{\sigma}^T(0)^2&=& \mu_0^2 + 3\lambda_0^2 \left[I_T +J_T +s_T^2\right]\nonumber\\
  c &=& s_T \left[\mu_0^{2} + \lambda_0^{2} s_T^{2}  +
  3 \lambda_0^{2} \left( I_T +J_T \right) \right] 
 \label{eq52}
 \end{eqnarray}
 \\
which is of identical form as the $T=0$ expressions.
 
We now move on to the RPA problem at finite $T$.   
In the spirit of the zero-temperature RPA an operator $Q_{\nu}^+$ 
is used which contains the same excitations as the symmetry generator. 
Therefore $Q_{\nu}^+$ is given by
\\
\begin{equation}
 Q_{\nu}^+ =
 X^1_{\nu} \alpha^{a\, +}_0 \,\, -\,\,  Y^1_{\nu} \alpha^a_0
 \,\,+\,\,
 \sum_{q}
  \left[ X_{\nu}^2(q)  \beta^+_q \alpha^{a\, +}_{-q}
 \, \,-\,\,  Y_{\nu}^2(q)  \beta_{-q} \alpha^{a}_q\,\,+\,\,
 X_{\nu}^3(q)  \beta_{-q} \alpha^{a\, +}_{-q}
 \, \,-\,\,  Y_{\nu}^3(q)  \beta^+_{q} \alpha^{a}_{q}
 \right]
\label{eq53}
\end{equation}
\\
which now contains additional terms of 'mixed' type.    
The analogous equations of motion have been worked out in 
refs.~\cite{Som83, TST85} and read
\\
\begin{equation}
 \ave { \left[ \delta Q_{\nu} \, \, , \, \, \left[H\, , \, Q_{\nu}^+
 \right] \right]  } = \Omega_{\nu}
 \ave{  \left[ \delta Q_{\nu}  \, , \, Q_{\nu}^+
 \right]  }
\label{eq54}
\end{equation}
\\
where the average is to be taken in the grand ensemble. 
The RPA operator (\ref{eq53}) now generates a set of six  equations 
for the various amplitudes which can be written as 
\\
\begin{equation}
\int d^3{\vec p}
 {\cal  M}_{ij}(q,p)  {\chi}_j(p) =
 \Omega^T_{\nu} {\cal N}_{ij}(q){\chi}_j(q)
\label{eq55}
\end{equation}
\\
where ${\cal M}(q,p)$ is the $6\times 6$ RPA matrix to be inverted, 
$\Omega^T_{\nu}$ and $\chi(p)$ are the eigenvalues and 6-column eigenvectors
respectively, and finally ${\cal N}(q)$ is the diagonal matrix norm :
\\
\begin{eqnarray}
{\cal N}_{11}(q) &=& - {\cal N}_{22}(q) = 1\nonumber\\
{\cal N}_{33}(q) &=& - {\cal N}_{44}(q) = 1+f_{\pi}(q)+f_{\sigma}(q)\nonumber\\
{\cal N}_{55}(q) &=& - {\cal N}_{66}(q) = f_{\sigma}(q)-f_{\pi}(q)\nonumber\\
\label{eq56}
\end{eqnarray}
\\
The normalization condition for the eigenvectors is \\
\begin{eqnarray}
< \left[ Q_{\nu} , Q^+_{\nu'} \right] > &\,=\,& \left(\mid X^1_{\nu} \mid^2 - \mid Y^1_{\nu} \mid^2\right) \,+\, \sum_{q}\left[\left[1+f_{\pi}(q)+f_{\sigma}(q)\right] \left( \mid X^2_{\nu}(q) \mid^2 - \mid Y^2_{\nu}(q) \mid^2\right)
\right. \nonumber\\
&\,+\,&  \left. \left[f_{\sigma}(q)-f_{\pi}(q)\right] \left( \mid X^3_{\nu}(q) \mid^2 - \mid Y^3_{\nu}(q) \mid^2\right)\right] \,=\, \delta_{\nu\nu'}
\label{eq57}
\end{eqnarray}
\\  
The solution of the eigenvalue problem in eq.(\ref{eq55}) gives\\
\\
\begin{equation}
\Omega_{\nu}^2 \,=\, {\cal E}^{T\, 2}_{\pi}(0) \quad +\quad  \frac{4\lambda_0^4 s_T^2 \Sigma^T_{\pi \sigma}(\Omega^2_{\nu})}{1-2\lambda_0^2 \Sigma_{\pi \sigma}^T(\Omega_{\nu}^2)}
\label{eq59} 
\end{equation}
\\
where 
\\
\begin{equation}
\Sigma_{\pi\sigma}^T(\Omega_{\nu}^2)\,=\,
\int \frac{d^3{\vec p}}{(2\pi)^3} \left[
\frac{{\cal E}^T_{\pi}(p)+{\cal E}^T_{\sigma}(p)}{2 {\cal E}^T_{\pi}(p){\cal E}^T_{\sigma}(p)} \frac{1+ f_{\pi}(p)+f_{\sigma}(p)}{
\Omega_{\nu}^2-({\cal E}^T_{\pi}(p)+{\cal E}^T_{\sigma}(p))^2}
\,+\, 
\frac{{\cal E}^T_{\pi}(p)-{\cal E}^T_{\sigma}(p)}{2 {\cal E}^T_{\pi}(p){\cal E}^T_{\sigma}(p)}
\frac{f_{\sigma}(p)-f_{\pi}(p)}{ 
\Omega_{\nu}^2-({\cal E}^T_{\pi}(p)-{\cal E}^T_{\sigma}(p))^2 }
\right]
\label{eq60} 
\end{equation}
\\
and the Bose occupation factors are given by \\
\\
\begin{equation}
f_{\pi, \sigma}(p) = \left[ exp(\frac{{\cal E}_{\pi, \sigma}^T(p)}{T}) - 1\right]^{-1}
\label{eq70} 
\end{equation}
One verifies that in the zero-temperature limit the previous HFB-RPA results
are recovered.
\\
For the physical interpretation of these results we first address the
question whether the FTHFB-RPA is able to describe the two realizations 
of the symmetry {\sl i.e.} the Wigner phase and the Goldstone phase.
In the Goldstone phase the vacuum condensate $s_T$ is finite and the bare 
mass $\mu_0^2$ negative such that the third equation in (\ref{eq52}) is 
satisfied in the chiral limit ($c=0$). There exists therefore a Goldstone 
mode in the theory and the symmetry is no longer manifest in the particle 
spectrum which means that the masses of the $\sigma$ and the 
$\pi$ are different. In the zero-temperature case we have demonstrated 
that HFB-RPA scheme is consistent with these requirements.\\ 

There is, however, an alternative solution of the third 
equation in (\ref{eq52}). Suppose the condensate $s_T$ vanishes
at some temperature. In this case the three-particle coupling disappears
from the interaction Hamiltonian, as can be seen explicitly from
eq.~(\ref{eq8}). The single-particle state no longer couples to two-particle 
states and the only contribution of the single-particle masses is the one 
that comes from the four-point interactions in the mean-field 
calculation. This can also be checked explicitly from the RPA eigenvalues. 
It is now easy to see from eqs.~(\ref{eq52}) that the masses become degenerate
{\sl i.e.} ${\cal E}_{\pi}={\cal E}_{\sigma}$ and we are in the Wigner mode.
The question is whether this is inconsistent with the fact that $Q^a_5$, the 
generator of the symmetry, commutes with the Hamiltonian which leads to a 
spurious solution of the RPA in the Goldstone phase. First one should note
that if the masses are equal then the thermal occupation factors for both the 
pion and the sigma are the same. This leads to 
${\cal N}_{55}(q) = {\cal N}_{66}(q) = 0$ implying that the RPA spurious mode 
cannot be normalized. Secondly, from equation (\ref{eq59}) we see that 
to have the zero frequency mode, one must fulfill the following condition\\
\\
\begin{equation}
 0 \,=\, {\cal E}^{T\, 2}_{\pi}(0)
 \,+\, \frac{4\lambda_0^4 s_T^2 \Sigma^T_{\pi \sigma}(0)}{1-2\lambda_0^2 
 \Sigma_{\pi \sigma}^T(0)}
\label{eq61} 
\end{equation}
\\
In analogy to eq.~(\ref{eq24}) one can prove the following identity
\\
\begin{equation}
 \Sigma^T_{\pi\sigma}(0) \,=\,
 \frac{I_T \,\, -\,\, J_T}
  { {\cal E}^{T\,2}_{\pi}(0)- {\cal E}^{T\,2}_{\sigma}(0)},
\label{eq62}
\end{equation}
\\
which is only true, however, if all six terms in the excitation operator
$Q^+_\nu$ (eq.~(\ref{eq53}) are kept. Now the condition for a spurious mode
solution can be simply recast as
\\
\begin{equation}
 \frac{c}{s_T}\,=\,0
\label{eq63}
\end{equation}
\\
which means that the ratio of the symmetry breaking term in the 
Lagrangian and the condensate must vanish to allow a zero-energy solution. 
This can only happen for finite $s_T$ {\sl i.e.} in the Goldstone phase. 
Once the Wigner phase is reached this condition can no longer be satisfied.
This reiterates the  fact that the spurious mode is a 
manifestation of a broken symmetry which disappears once the latter is restored.

\section{Conclusions and outlook}
In summary we have presented a non-perturbative method
based on the well-known selfconsistent QRPA formalism for studying
the linear $\sigma$-model in the bosonic sector. Being 'symmetry
conserving' the method yields a zero mode in the chiral limit. This
is required by Goldstone's theorem for a spontaneously broken symmetry.
While, at field level, the pion acquires a mass through the BCS
mechanism irrespective of the explicit symmetry breaking term in the 
Lagrangian the inclusion of RPA correlations removes this artifact.   
We have also demonstrated that the extension of the QRPA to finite 
temperature is workable and reproduces the expected result that the 
zero mode persists at finite temperature. Applications 
to the $SU(3)$-case as well as the inclusion of fermions are 
straightforward and are being considered. This will hopefully
provide new insight into the nature of the chiral phase transition.
Since, in contrast to Nambu-Jona-Lasinio type models, the linear 
$\sigma$-model is renormalizable a program of non-perturbative 
renormalisation should be persued in order to assess its impact on
the physics. Despite the selfconsistency of the mean-field equations 
a solution of this problem does not seem out of reach.
A challenging problem is the application of the 
formalism to the two-pion case. Here one attempts to build a scattering
equation which is consistent with the low-energy theorems required
by the symmetry. As is known from the analogous fermionic problem
higher RPA schemes have to be employed. In particular the second RPA
\cite{DNSW, YDG83} is also 'symmerty conserving'. Its bosonic analog is easy to
construct but some conceptual problems remain to be resolved.

\end{document}